\documentclass[11pt]{cernrep}
\usepackage{graphicx}
\usepackage{here}
\begin{document}


\thispagestyle{empty}

\onecolumn
\begin{flushright}
{\large \begin{tabular}{l}
DESY 02--102\\
hep--ph/0207324\\
July 2002
\end{tabular}}
\end{flushright}

\vspace*{3.5truecm}

\begin{center}
\boldmath
{\LARGE \bf Extracting $\gamma$ Through Flavour-Symmetry Strategies}
\unboldmath

\vspace*{3.2cm}

{\sc{\Large Robert Fleischer}}

\vspace*{0.1truecm}

{\it{\large Deutsches Elektronen-Synchrotron DESY, 
Notkestr.\ 85, D--22607 Hamburg, Germany}}

\vspace{3.2truecm}

{\large\bf Abstract\\[10pt]} \parbox[t]{\textwidth}{
A brief overview of flavour-symmetry strategies to extract the angle $\gamma$ 
of the unitarity triangle is given, focusing on $B\to\pi K$ modes and the 
$B_d\to\pi^+\pi^-$, $B_s\to K^+K^-$ system. We discuss also a variant of the 
latter approach for the $e^+e^-$ $B$-factories, where $B_s\to K^+K^-$ is 
replaced by $B_d\to\pi^\mp K^\pm$.
}

\vspace{3.5cm}
 
{\sl Contribution to the Proceedings of the\\ 
Workshop on the CKM Unitarity Triangle,\\ 
CERN, Geneva, 13--16 February 2002}

\end{center}

\newpage
\thispagestyle{empty}
\vbox{}
\newpage
 
\setcounter{page}{1}
 

\boldmath
 \title{EXTRACTING $\gamma$ THROUGH FLAVOUR-SYMMETRY STRATEGIES}
\unboldmath
\author{R. Fleischer}
\institute{Deutsches Elektronen-Synchrotron DESY,
Notkestra\ss e 85, D--22607 Hamburg, Germany}
\maketitle
\begin{abstract}
A brief overview of flavour-symmetry strategies to extract the angle $\gamma$ 
of the unitarity triangle is given, focusing on $B\to\pi K$ modes and the 
$B_d\to\pi^+\pi^-$, $B_s\to K^+K^-$ system. We discuss also a variant of the 
latter approach for the $e^+e^-$ $B$-factories, where $B_s\to K^+K^-$ is 
replaced by $B_d\to\pi^\mp K^\pm$. 
\end{abstract}

\section{INTRODUCTION}
An important element in the testing of the Kobayashi--Maskawa picture of
CP violation is the direct determination of the angle $\gamma$ of the 
unitarity triangle of the CKM matrix. Here the goal is to overconstrain 
this angle as much as possible. In the presence of new physics, discrepancies 
may arise between different strategies, as well as with the ``indirect'' 
results for $\gamma$ that are provided by the usual fits of the unitarity 
triangle, yielding at present $\gamma\sim 60^\circ$ \cite{UT-fits}. 

There are many approaches on the market to determine $\gamma$ (for
a detailed review, see Ref.~\cite{RF-Phys-Rep}). Here we shall focus on
$B\to\pi K$ modes \cite{GRL}--\cite{GR-BpiK-recent}, which can be analysed 
through flavour-symmetry arguments and plausible dynamical assumptions, and 
the $U$-spin-related decays $B_d\to \pi^+\pi^-$, $B_s\to K^+K^-$
\cite{RF-BsKK}. The corresponding flavour-symmetry strategies allow the 
determination of $\gamma$ and valuable hadronic parameters with a ``minimal'' 
theoretical input. Alternative approaches, relying on a more extensive use of 
theory, are provided by the recently developed ``QCD factorization'' 
\cite{QCDF} and ``PQCD'' \cite{PQCD} approaches, which allow furthermore a 
reduction of the theoretical uncertainties of the flavour-symmetry strategies 
discussed here. Let us note that these approaches are also particularly 
promising from a practical point of view: BaBar, Belle and CLEO-III may 
probe $\gamma$ through $B\to\pi K$ modes, whereas the $U$-spin strategy, 
requiring also a measurement of the $B_s$-meson decay $B_s\to K^+K^-$, is 
already interesting for run II of the Tevatron \cite{TEV-II}, and can be 
fully exploited in the LHC era \cite{LHC-Report}. A variant for the 
$B$-factories \cite{U-variant}, where $B_s\to K^+K^-$ is replaced by 
$B_d\to\pi^\mp K^\pm$, points already to an exciting picture \cite{FlMa2}.

\boldmath
\section{$B\to \pi K$ DECAYS}
\unboldmath
Using the isospin flavour symmetry of strong interactions, relations between 
$B\to\pi K$ amplitudes can be derived, which suggest the following 
combinations to probe $\gamma$: the ``mixed'' $B^\pm\to\pi^\pm K$, 
$B_d\to\pi^\mp K^\pm$ 
system \cite{PAPIII}--\cite{defan}, the ``charged'' $B^\pm\to\pi^\pm K$, 
$B^\pm\to\pi^0K^\pm$ system \cite{NR}--\cite{BF-neutral1}, and the 
``neutral'' $B_d\to\pi^0 K$, $B_d\to\pi^\mp K^\pm$ system 
\cite{BF-neutral1,BF-neutral2}.
Interestingly, already CP-averaged $B\to\pi K$ branching ratios
may lead to non-trivial constraints on $\gamma$ \cite{FM,NR}. In order
to {\it determine} this angle, also CP-violating rate differences have
to be measured. To this end, we introduce the following observables
\cite{BF-neutral1}:
\begin{equation}\label{mixed-obs}
\left\{\begin{array}{c}R\\A_0\end{array}\right\}
\equiv\left[\frac{\mbox{BR}(B^0_d\to\pi^-K^+)\pm
\mbox{BR}(\overline{B^0_d}\to\pi^+K^-)}{\mbox{BR}(B^+\to\pi^+K^0)+
\mbox{BR}(B^-\to\pi^-\overline{K^0})}\right]\frac{\tau_{B^+}}{\tau_{B^0_d}}
\end{equation}
\begin{equation}\label{charged-obs}
\left\{\begin{array}{c}R_{\rm c}\\A_0^{\rm c}\end{array}\right\}
\equiv2\left[\frac{\mbox{BR}(B^+\to\pi^0K^+)\pm
\mbox{BR}(B^-\to\pi^0K^-)}{\mbox{BR}(B^+\to\pi^+K^0)+
\mbox{BR}(B^-\to\pi^-\overline{K^0})}\right]
\end{equation}
\begin{equation}\label{neut-obs}
\left\{\begin{array}{c}R_{\rm n}\\A_0^{\rm n}\end{array}\right\}
\equiv\frac{1}{2}\left[\frac{\mbox{BR}(B^0_d\to\pi^-K^+)\pm
\mbox{BR}(\overline{B^0_d}\to\pi^+K^-)}{\mbox{BR}(B^0_d\to\pi^0K^0)+
\mbox{BR}(\overline{B^0_d}\to\pi^0\overline{K^0})}\right].
\end{equation}

If we employ the isospin flavour symmetry and make plausible dynamical 
assumptions, concerning mainly the smallness of certain rescattering 
processes, we obtain parametrizations of the following structure
\cite{defan,BF-neutral1} (for alternative ones, see Ref.~\cite{neubert}):
\begin{equation}\label{obs-par}
R_{({\rm c,n})},\, A_0^{({\rm c,n})}=
\mbox{functions}\left(q_{({\rm c,n})}, r_{({\rm c,n})},
\delta_{({\rm c,n})}, \gamma\right).
\end{equation}
Here $q_{({\rm c,n})}$ denotes the ratio of electroweak (EW) penguins to 
``trees'', $r_{({\rm c,n})}$ is the ratio of ``trees'' to QCD penguins, and
$\delta_{({\rm c,n})}$ the strong phase between ``trees'' and QCD penguins. 
The EW penguin parameters $q_{({\rm c,n})}$ can be fixed through theoretical 
arguments: in the mixed system \cite{PAPIII}--\cite{GR}, we have $q\approx0$, 
as EW penguins contribute only in colour-suppressed form; in the charged and 
neutral $B\to\pi K$ systems, $q_{\rm c}$ and $q_{\rm n}$ can be fixed through 
the $SU(3)$ flavour symmetry without dynamical assumptions 
\cite{NR}--\cite{BF-neutral2}. The $r_{({\rm c,n})}$ can be determined with 
the help of additional experimental information: in the mixed system, $r$ 
can be fixed through arguments based on factorization \cite{PAPIII,GR,QCDF} 
or $U$-spin \cite{GR-Uspin}, whereas $r_{\rm c}$ and $r_{\rm n}$ can be 
determined from the CP-averaged $B^\pm\to\pi^\pm\pi^0$ branching ratio by 
using only the $SU(3)$ flavour symmetry \cite{GRL,NR}. The uncertainties 
arising in this programme from $SU(3)$-breaking effects can be reduced 
through the QCD factorization approach \cite{QCDF}, which is moreover in 
favour of small rescattering processes. For simplicity, we shall neglect 
such FSI effects in the discussion given below. 

Since we are in a position to fix the parameters $q_{({\rm c,n})}$ 
and $r_{({\rm c,n})}$, we may determine $\delta_{({\rm c,n})}$ 
and $\gamma$ from the observables given in (\ref{obs-par}). This can 
be done separately for the mixed, charged and neutral $B\to\pi K$ 
systems. It should be emphasized that also CP-violating rate 
differences have to be measured to this end. Using 
just the CP-conserving observables $R_{({\rm c,n})}$, we may obtain 
interesting constraints on $\gamma$. In contrast to $q_{({\rm c,n})}$ 
and $r_{({\rm c,n})}$, the strong phase $\delta_{({\rm c,n})}$ suffers 
from large hadronic uncertainties. However, we can get rid of 
$\delta_{({\rm c,n})}$ by keeping it as a ``free'' variable, yielding 
minimal and maximal values for $R_{({\rm c,n})}$:
\begin{equation}\label{const1}
\left.R^{\rm ext}_{({\rm c,n})}\right|_{\delta_{({\rm c,n})}}=
\mbox{function}\left(q_{({\rm c,n})},r_{({\rm c,n})},\gamma\right).
\end{equation}
Keeping in addition $r_{({\rm c,n})}$ as a free variable, we obtain 
another -- less restrictive -- minimal value 
\begin{equation}\label{const2}
\left.R^{\rm min}_{({\rm c,n})}\right|_{r_{({\rm c,n})},\delta_{({\rm c,n})}}
=\mbox{function}\left(q_{({\rm c,n})},\gamma\right)\sin^2\gamma.
\end{equation}
These extremal values of $R_{({\rm c,n})}$ imply 
constraints on $\gamma$, since the cases corresponding to
$R^{\rm exp}_{({\rm c,n})}< R^{\rm min}_{({\rm c,n})}$
and $R^{\rm exp}_{({\rm c,n})}> R^{\rm max}_{({\rm c,n})}$
are excluded. Present experimental data seem to point towards values for
$\gamma$ that are {\it larger} than $90^\circ$, which would be in conflict
with the CKM fits, favouring $\gamma\sim60^\circ$ \cite{UT-fits}. 
Unfortunately, the present experimental uncertainties do not yet allow us 
to draw definite conclusions, but the picture should improve significantly 
in the future. 

An efficient way to represent the situation in the $B\to\pi K$ system 
is provided by allowed regions in the $R_{({\rm c,n})}$--$A_0^{({\rm c,n})}$
planes \cite{FlMa1,FlMa2}, which can be derived within the Standard Model
and allow a direct comparison with the experimental data. A complementary
analysis in terms of $\gamma$ and $\delta_{\rm c,n}$ was performed in 
Ref.~\cite{ital-corr}. Another recent $B\to\pi K$ study can be found in 
Ref.~\cite{GR-BpiK-recent}, where the $R_{\rm (c)}$ were calculated for given 
values of $A_0^{\rm (c)}$ as functions of $\gamma$, and were compared with 
the $B$-factory data. In order to analyse $B\to\pi K$ modes, also certain
sum rules may be useful \cite{matias}.

\boldmath
\section{THE $B_d\to\pi^+\pi^-$, $B_s\to K^+K^-$ SYSTEM}
\unboldmath
As can be seen from the corresponding Feynman diagrams, $B_s\to K^+K^-$ 
is related to $B_d\to\pi^+\pi^-$ through an interchange of all down and 
strange quarks. The decay amplitudes read as follows \cite{RF-BsKK}:
\begin{equation}\label{Bdpipi-ampl0}
A(B_d^0\to\pi^+\pi^-)\propto\left[e^{i\gamma}-d e^{i\theta}\right],\quad
A(B_s^0\to K^+K^-)\propto
\left[e^{i\gamma}+\left(\frac{1-\lambda^2}{\lambda^2}\right)
d'e^{i\theta'}\right],
\end{equation}
where the CP-conserving strong amplitudes $d e^{i\theta}$ and 
$d'e^{i\theta'}$ measure, sloppily speaking, ratios of penguin to tree 
amplitudes in $B_d^0\to\pi^+\pi^-$ and $B_s^0\to K^+K^-$, respectively. 
Using these general parametrizations, we obtain expressions for the 
direct and mixing-induced CP asymmetries of the following kind:
\begin{equation}\label{Bpipi-obs}
{\cal A}_{\rm CP}^{\rm dir}(B_d\to\pi^+\pi^-)=
\mbox{function}(d,\theta,\gamma),\,
{\cal A}_{\rm CP}^{\rm mix}(B_d\to\pi^+\pi^-)=
\mbox{function}(d,\theta,\gamma,\phi_d=2\beta)
\end{equation}
\begin{equation}\label{BsKK-obs}
{\cal A}_{\rm CP}^{\rm dir}(B_s\to K^+K^-)=
\mbox{function}(d',\theta',\gamma),\,
{\cal A}_{\rm CP}^{\rm mix}(B_s\to K^+K^-)=
\mbox{function}(d',\theta',\gamma,\phi_s\approx0).
\end{equation}

Consequently, we have four observables at our disposal, depending on six 
``unknowns''. However, since $B_d\to\pi^+\pi^-$ and $B_s\to K^+K^-$ are 
related to each other by interchanging all down and strange quarks, the 
$U$-spin flavour symmetry of strong interactions implies
\begin{equation}\label{U-spin-rel}
d'e^{i\theta'}=d\,e^{i\theta}.
\end{equation}
Using this relation, the four observables in (\ref{Bpipi-obs},\ref{BsKK-obs}) 
depend on the four quantities $d$, $\theta$, $\phi_d=2\beta$ and $\gamma$, 
which can hence be determined \cite{RF-BsKK}. The theoretical accuracy is
only limited by the $U$-spin symmetry, as no dynamical assumptions about
rescattering processes have to be made. Theoretical considerations give us
confidence into (\ref{U-spin-rel}), as it does not receive $U$-spin-breaking 
corrections in factorization \cite{RF-BsKK}. Moreover, we may also obtain 
experimental insights into $U$-spin breaking \cite{RF-BsKK,gronau-U-spin}. 

The $U$-spin arguments can be minimized, if the $B^0_d$--$\overline{B^0_d}$ 
mixing phase $\phi_d=2\beta$, which can be fixed through 
$B_d\to J/\psi K_{\rm S}$, is used as an input. The observables 
${\cal A}_{\rm CP}^{\rm dir}(B_d\to\pi^+\pi^-)$ and
${\cal A}_{\rm CP}^{\rm mix}(B_d\to\pi^+\pi^-)$ allow us then to 
eliminate the strong phase $\theta$ and to determine $d$ as a function of
$\gamma$. Analogously, ${\cal A}_{\rm CP}^{\rm dir}(B_s\to K^+K^-)$ and 
${\cal A}_{\rm CP}^{\rm mix}(B_s\to K^+K^-)$ allow us to eliminate 
the strong phase $\theta'$ and to determine $d'$ as a function of
$\gamma$. The corresponding contours in the $\gamma$--$d$
and $\gamma$--$d'$ planes can be fixed in a {\it theoretically clean} way.
Using now the $U$-spin relation $d'=d$, these contours allow the 
determination both of the CKM angle $\gamma$ and of the hadronic quantities 
$d$, $\theta$, $\theta'$; for a detailed illustration, see Ref.~\cite{RF-BsKK}.
This approach is very promising for run II of the Tevatron and the experiments
of the LHC era, where experimental accuracies for $\gamma$ of 
${\cal O}(10^\circ)$ \cite{TEV-II} and ${\cal O}(1^\circ)$ \cite{LHC-Report} 
may be achieved, respectively. It should be emphasized that not only 
$\gamma$, but also the hadronic parameters $d$, $\theta$, $\theta'$ are of 
particular interest, as they can be compared with theoretical predictions, 
thereby allowing valuable insights into hadron dynamics. For other recently 
developed $U$-spin strategies, the reader is referred to 
Refs.~\cite{GR-Uspin,U-spin-other}.

\boldmath
\section{THE $B_d\to\pi^+\pi^-$, $B_d\to \pi^\mp K^\pm$ SYSTEM AND IMPLICATIONS
FOR $B_s\to K^+K^-$}
\unboldmath
A variant of the $B_d\to \pi^+\pi^-$, $B_s\to K^+K^-$ approach was developed
for the $e^+e^-$ $B$-factories \cite{U-variant}, where $B_s\to K^+K^-$ is not 
accessible: as $B_s\to K^+K^-$ and $B_d\to\pi^\mp K^\pm$ are related to each 
other through an interchange of the $s$ and $d$ spectator quarks, we may 
replace the $B_s$ mode approximately through its $B_d$ counterpart, which has 
already been observed by BaBar, Belle and CLEO. Following these lines and 
using experimental information on the CP-averaged $B_d\to\pi^\mp K^\pm$ and 
$B_d\to\pi^+\pi^-$ branching ratios, the relevant hadronic penguin parameters 
can be constrained, implying certain allowed regions in observable space
\cite{FlMa2}. An interesting situation arises now in view of the recent 
$B$-factory measurements of CP violation in $B_d\to\pi^+\pi^-$, allowing us 
to obtain new constraints on $\gamma$ as a function of the 
$B^0_d$--$\overline{B^0_d}$ mixing phase $\phi_d$, which is fixed through 
${\cal A}_{\rm CP}^{\rm mix}(B_d\to J/\psi K_{\rm S})$ up to a twofold 
ambiguity, $\phi_d\sim 51^\circ$ or $129^\circ$.
If we assume that ${\cal A}_{\rm CP}^{\rm mix}(B_d\to \pi^+\pi^-)$ is 
positive, as indicated by recent Belle data, and that $\phi_d$ is in 
agreement with the ``indirect'' fits of the unitarity triangle, i.e.\
$\phi_d\sim 51^\circ$, also the corresponding values for $\gamma$ 
around $60^\circ$ can be accommodated. On the other hand, for the second 
solution $\phi_d\sim129^\circ$, we obtain a gap around $\gamma\sim60^\circ$,
and could easily accommodate values for $\gamma$ larger than $90^\circ$.  
Because of the connection between the two solutions for $\phi_d$ and the 
resulting values for $\gamma$, it is very desirable to resolve the twofold 
ambiguity in the extraction of $\phi_d$ directly. As far as $B_s\to K^+K^-$ 
is concerned, the data on the CP-averaged $B_d\to \pi^+\pi^-$, 
$B_d\to\pi^\mp K^\pm$ branching ratios imply a very constrained allowed 
region in the space of ${\cal A}_{\rm CP}^{\rm mix}(B_s\to K^+K^-)$ and 
${\cal A}_{\rm CP}^{\rm dir}(B_s\to K^+K^-)$ within the Standard Model, 
thereby providing a narrow target range for run II of the Tevatron and 
the experiments of the LHC era \cite{FlMa2}. Other recent studies related to
$B_d\to\pi^+\pi^-$ can be found in Refs.~\cite{GR-BpiK-recent,Bpipi-recent}.

\section*{ACKNOWLEDGEMENTS}
I would like to thank Andrzej Buras, Thomas Mannel and Joaquim Matias for
pleasant collaborations on the topics discussed above.


\begin{thebibliography}{99}
\bibitem{UT-fits}See, for instance, A.~J.~Buras, F.~Parodi and A.~Stocchi,
TUM-HEP-465-02 [hep-ph/0207101];\\
A.~H\"ocker, H.~Lacker, S.~Laplace and F.~Le Diberder,
{Eur.\ Phys.\ J.}\ C {\bf 21} (2001) 225;\\
M.~Ciuchini {\it et al.},
{JHEP} {\bf 0107} (2001) 013.

\bibitem{RF-Phys-Rep}R.~Fleischer, DESY-THESIS-2002-022 [hep-ph/0207108],
to appear in Physics Reports.

\bibitem{GRL}M.~Gronau, J.~L.~Rosner and D.~London,
{Phys.\ Rev.\ Lett.}\  {\bf 73} (1994) 21.

\bibitem{PAPIII}R.~Fleischer,
{ Phys.\ Lett.}\ B {\bf 365} (1996) 399.

\bibitem{FM}R.~Fleischer and T.~Mannel,
{Phys.\ Rev.}\ D {\bf 57} (1998) 2752.

\bibitem{GR}M.~Gronau and J.~L.~Rosner,
{Phys.\ Rev.}\ D {\bf 57} (1998) 6843.

\bibitem{defan}R.~Fleischer,
{Eur.\ Phys.\ J.}\ C {\bf 6} (1999) 451.

\bibitem{NR}M.~Neubert and J.~L.~Rosner,
{Phys.\ Lett.}\ B {\bf 441} (1998) 403;\\
{Phys.\ Rev.\ Lett.}\  {\bf 81} (1998) 5076.

\bibitem{neubert}M.~Neubert,
{JHEP} {\bf 9902} (1999) 014.

\bibitem{BF-neutral1}A.~J.~Buras and R.~Fleischer,
{Eur.\ Phys.\ J.}\ C {\bf 11} (1999) 93.

\bibitem{BF-neutral2}A.~J.~Buras and R.~Fleischer,
{Eur.\ Phys.\ J.}\ C {\bf 16} (2000) 97.

\bibitem{FlMa1}R.~Fleischer and J.~Matias,
Phys.\ Rev.\ D {\bf 61} (2000) 074004.

\bibitem{ital-corr}M.~Bargiotti {\it et al.}, 
Eur.\ Phys.\ J.\ C {\bf 24} (2002) 361.

\bibitem{GR-BpiK-recent}M.~Gronau and J.~L.~Rosner,
{Phys.\ Rev.}~{\bf D65} (2002) 013004 [E: {\bf D65} (2002) 079901].

\bibitem{RF-BsKK}R.~Fleischer,
{Phys.\ Lett.}\ B {\bf 459} (1999) 306.

\bibitem{QCDF}M.~Beneke, 
G.~Buchalla, M.~Neubert and C.~T.~Sachrajda,
{Phys.\ Rev.\ Lett.}\  {\bf 83} (1999) 1914;\\
{Nucl.\ Phys.}\ B {\bf 606} (2001) 245.

\bibitem{PQCD}H.-n.~Li and H.~L.~Yu,
{Phys.\ Rev.}\ D {\bf 53} (1996) 2480;\\
Y.~Y.~Keum, H.-n.~Li and A.~I.~Sanda,
{Phys.\ Lett.}\ B {\bf 504} (2001) 6;\\
Y.~Y.~Keum and H.-n.~Li,
{Phys.\ Rev.}\ D {\bf 63} (2001) 074006.

\bibitem{TEV-II}K.~Anikeev {\it et al.}, FERMILAB-Pub-01/197 [hep-ph/0201071].

\bibitem{LHC-Report}P.~Ball {\it et al.}, CERN-TH-2000-101
[hep-ph/0003238].

\bibitem{U-variant}R.~Fleischer,
{Eur.\ Phys.\ J.}\ C {\bf 16} (2000) 87.

\bibitem{FlMa2}R.~Fleischer and J.~Matias,
DESY-02-040 [hep-ph/0204101], to appear in Phys.\ Rev.\ D.

\bibitem{GR-Uspin}
M.~Gronau and J.~L.~Rosner,
{Phys.\ Lett.}\ B {\bf 482} (2000) 71.

\bibitem{matias}J.~Matias,
Phys.\ Lett.\ B {\bf 520} (2001) 131.

\bibitem{gronau-U-spin}M.~Gronau,
{Phys.\ Lett.}\ B {\bf 492} (2000) 297.

\bibitem{U-spin-other}R.~Fleischer,
{Eur.\ Phys.\ J.}\ C {\bf 10} (1999) 299,
Phys.\ Rev.\ D {\bf 60} (1999) 073008;\\
P.~Z.~Skands,
{JHEP} {\bf 0101} (2001) 008.

\bibitem{Bpipi-recent}
M.~Gronau and J.~L.~Rosner,
Phys.\ Rev.\ D {\bf 65} (2002) 093012, 
113008 
and\\ 
TECHNION-PH-2002-21 [hep-ph/0205323];\\
C.-D.~L\"u and Z.-j.~Xiao,
BIHEP-TH-2002-22 [hep-ph/0205134].

\end{thebibliography}
\end{document}